\documentclass[aps,prd,twocolumn,superscriptaddress,amsmath,amssymb]{revtex4-1}
\usepackage{url}
\usepackage{hyperref}
\usepackage{mathtools}
\usepackage{tensor}
\usepackage{palatino}
\usepackage{tikz}
\usepackage{multirow}
\usepackage{tabularx}
\usetikzlibrary{shapes.geometric, arrows}

\begin{document}

\title{Inflation in the general Poincar\'e gauge cosmology}

\author{Hongchao Zhang}
\email{zhanghc@mail.dlut.edu.cn}
\affiliation{Institute of Theoretical Physics, School of Physics, \\ Dalian University of Technology, Dalian, 116024, P. R. China}
\affiliation{Institute for Gravitation and the Cosmos, \\ Penn State University, University Park, PA 16801, U.S.A.}
\author{Lixin Xu}
\email{lxxu@dlut.edu.cn}
\affiliation{Institute of Theoretical Physics, School of Physics, \\ Dalian University of Technology, Dalian, 116024, P. R. China}

\date{\today}

\begin{abstract}
The general Poincar\'e gauge cosmology given by a nine-parameter gravitational Lagrangian with ghost- and tachyon-free conditions is studied from the perspective of field theory. By introducing new variables for replacing two (pseudo-) scalar torsions, the Poincar\'e gauge cosmological system can be recast into a gravitational system coupled to two-scalar fields with a potential up to quartic-order. We discussed the possibility of this system producing two types of inflation without any extra inflatons. The hybrid inflation with a first-order phase transition can be ruled out, while the slow rollover can be achieved. The numerical analysis shows that the two-scalar fields system evolved in a potential well processes spontaneously four stages: ``pre-inflation'', slow-roll inflation with large enough e-folds, ``pre-reheating'' and reheating. We also studied the stableness of this system by setting large values of initial kinetic energies. The results show that even if the system evolves past the highest point of the potential well, the scalar fields can still return to the potential well and cause inflation. The general Poincar\'e gauge cosmology provides us with a self-consistent candidate of inflation.
\end{abstract}


\maketitle

\section{Introduction}
The standard model (SM) framework of cosmology based on Einstein's general relativity (GR) is quite successful in describing the evolution of the Universe on large enough scales \cite{aghanim2018planck}.
The SM infers that the Universe experienced a significant accelerated expansion in the very early period, which is called the inflation.
By introducing the inflation, problems such as the horizon, the flatness, the origin of perturbations, and the monopoles, that have plagued cosmologists before 1980s, can be solved naturally \cite{guth1981inflationary}.
After years of development, some inflationary models, such as the standard single-field (inflaton) and Starobinsky's inflation, can match the current observations in very high precision \cite{akrami2018planck,martin2014best}.
Unfortunately, those models lack a more essential mechanism for the origin of the inflaton(s), namely the source(s) of inflation.
The classical theory of inflation requires the Universe to experience an exponentially accelerated expansion from about $10^{-36}s$ to $10^{-32}s$ in the cosmic chronology \cite{garcia2005cosmology}.
This expansion drove the spatial curvature of the Universe towards extreme flatness, and established the causal correlations on the uniformity of the cosmic microwave background (CMB), and generated the seeds of large-scale structure \cite{baumann2009tasi}.
At the end of this stage, the expansion decelerated spontaneously and the Universe exited from the adiabatic process.
The subsequent reheating led to various particles to be generated \cite{kofman1997towards,bassett2006inflation}.
According to the way of exiting the expansion, the inflationary models can be classified into the slow rollover and the first-order phase transition \cite{linde1994hybrid}.
In addition, an effective correction from the loop quantum cosmology (LQC) enables to push the beginning of the whole process back to the Planck scale, where the big bang singularity have been replaced by the big bounce \cite{agullo2012quantum,agullo2013pre}.
The mechanism of inflation from big bounce to reheating is clear phenomenologically.

As a single-field model, Starobinsky's inflation given by a Lagrangian $\tilde{R}+\tilde{R}^2/6M^2$ plus some small non-local terms (which are crucial for reheating after inflation) is an internally self-consistent cosmological model, which possess a (quasi-)de Sitter stage in the early Universe with slow-roll decay, and a graceful exit to the subsequent radiation-dominated Friedmann-Lema\^{i}tre-Robertson-Walker (FLRW) stage \cite{starobinsky1980new,vilenkin1985classical,mijic1986r}.
This is one of the most appealing from both theoretical and observational perspectives among different models of inflation \cite{castellanos2018higher}.

Besides adding directly higher order curvature invariants or scalar fields to the Einstein-Hilbert (EH) action, another more fundamental way to generalize GR from the geometric and gauge perspectives has been introduced systematically since 1970's \cite{hehl1976general,blagojevic2013gauge}, which is called the Poincar\'e gauge gravity (PGG).
As the maximum group of Minkowski spacetime isometrics, Poincar\'e group possesses both translations and rotations, which totally has $10$ degree of freedom.
If constructing a gauge field theory based on the local invariance of the Poincar\'e group, the gravity will be represented by two independent gauge fields: tetrads $e$ and spin-connections $\omega$, corresponding to the translations and rotations, respectively.
Analogous to the Yang-Mills theory, one can verify that torsion $T$ and curvature $R$ are just their gauge field strengths.
According to the Noether's theorems, the symmetries of translation and rotation lead to two conservation objects: energy-momentum and spin spin-angular momentum.
Further more, the energy-momentum can be connected through Einstein's equation with curvature, and the spin-angular momentum with torsion through Cartan's equation, which mean that the sources of spacetime curvature and torsion are energy-momentum and spin of matter, respectively.
The above are fundamental ideas of PGG, which follows the schemes of standard Yang-Mills theory.
From the geometrical perspective, the spacetime extends from Riemann's to Riemann-Cartan's, where curvature measures the difference of a vector after parallel transporting along a infinitesimal loop, and torsion for the failure of closure of the parallelogram made of the infinitesimal displacement.
In order to show the extension of PGG to GR, we plot the following diagram:
\tikzstyle{Einstein} = [rectangle, rounded corners, minimum width = 1.cm, minimum height=0.5cm,text centered,text width=2cm, draw = blue]
\tikzstyle{Cartan} = [rectangle, rounded corners, minimum width = 1.cm, minimum height=0.5cm,text centered,text width=2.5cm, draw = red]
\tikzstyle{EinLor} = [rectangle, minimum width=0.5cm, minimum height=0.5cm, text centered,text width=2cm, draw=blue]
\tikzstyle{CarPoi} = [rectangle, minimum width=0.5cm, minimum height=0.5cm, text centered,text width=2cm, draw=red]
\tikzstyle{arrow} = [thin,<->,>=stealth]
\tikzstyle{arrow0} = []
\tikzstyle{arrow1} = [thin,->,>=stealth]
\tikzstyle{arrow2} = [thin,->,>=stealth]
\begin{center}
	\begin{tikzpicture}[node distance=2cm]
	\node[Einstein](cur){Curvature};
	\node[Einstein, right of = cur, xshift = 1.0cm, yshift = -2.7cm](con){Connection};
	\node[Einstein, left of = cur, xshift = -1.0cm, yshift = -2.7cm](EMT){Energy-Momentum};
	\node[Cartan, right of = EMT, xshift = 0.01cm, yshift = -2.7cm, inner sep=0.001pt](tra){Translation};
	\node[Einstein, left of = con, xshift = -0.01cm, yshift = -2.7cm, inner sep=0.001pt](rot){Rotation};
	\node[Cartan, left of = tra, xshift = -0.01cm, yshift = -2.7cm](can){Canonical $1$-form};
	\node[Cartan, right of = rot, xshift = 0.01cm, yshift = -2.7cm](SAM){Spin Angular Momentum};
	\node[Cartan, right of = can, xshift = 1.0cm, yshift = -2.7cm](tor){torsion};
	\node[CarPoi, left of = tra, xshift = -0.5cm, inner sep=0.001pt](PG){Poincar\'e};
	\node[EinLor, right of = rot, xshift = 0.5cm, inner sep=0.001pt](LG){Lorentz};
	\node[CarPoi, above of = tor, yshift = 0.5cm](Car){Cartan};
	\node[EinLor, below of = cur, yshift = -0.5cm](Ein){Einstein};
	\coordinate (point1) at (-3cm, -6cm);
	\draw [arrow] (cur) -- node [above,rotate=-35] {$2^{nd}$ structure} node [below,rotate=-35] {$R=\mathcal{D}_\omega \omega$} (con.north);
	\draw [arrow] (cur) -- node [above,rotate=35] {Einstein eq.} (EMT.north);
	\draw [arrow] (EMT.south) -- node [above,rotate=-45] {Noether's thm.} (tra.north);
	\draw [arrow] (rot.north) -- node [above,rotate=45] {Gauge potential} (con.south);
	\draw [arrow] (tra.south) -- node [below,rotate=45] {Gauge potential} (can.north);
	\draw [arrow] (can.south) -- node [above,rotate=-35] {$T=\mathcal{D}_\omega \theta$} node [below,rotate=-35] {$1^{st}$ structure} (tor);
	\draw [arrow0] (rot) -- node [] {$\oplus$} (tra);
	\draw [arrow] (tor) -- node [below,rotate=35] {Cartan eq.} (SAM.south);
	\draw [arrow] (SAM.north) -- node [below,rotate=-45] {Noether's thm.} (rot.south);
	\draw [arrow1] (tra) -- (PG);
	\draw [arrow2] (LG) -- (rot);
	\end{tikzpicture}
\end{center}
General speaking, the crucial different between PGG and GR-based theories, such as $f(R)$ gravity, is that the former removed the restriction of torsion-free.
However, the direct generalization from the EH action will be back to Einstein's theory, when the spin tensor of matter vanishes because of the algebraic Cartan equation, i.e. torsion can not propagate.
This reminds us that in order to obtain the propagating torsion in the vacuum, the action should be also generalized.
The standard PGG Lagrangian has a quadratic field strength form \cite{nester2017gravity}:
\begin{equation}
\mathcal{L}_G\sim \Lambda+curvature+torsion^2+\frac{1}{\varrho}curvature^2,
\end{equation}
where $\Lambda$ is the cosmological constant, and $\varrho$ the parameter with certain dimension.
The additional quadratic terms are naturally at most second derivative if one regards tetrads and spin-connections as the fundamental variables.
It is likely that such terms introduce ghost degrees of freedom, when one considers the particle substance of the gravity.
That would be something troublesome even for a simple modified gravity theory.
The existence of the ghost is closely related to the fact that the modified equation of motion has orders of time-derivative higher than two, for example, scale factor $a$ will be fourth-order over time in the general quadratic curvature case in FLRW cosmology.
Due to Ostrogradsky's theorem \cite{ostrogradsky1850memoires}, a system is not (kinematically) stable if it is described by a non-degenerate higher time-derivative Lagrangian.
To avoid the ghosts, a bunch of scalar-tensor theories of gravity was introduced, such as the Horndeski theory and beyond \cite{kobayashi2019horndeski,langlois2016degenerate}.
Another way to evade Ostrogradsky's theorem is to break Lorentz invariance in the ultraviolet and include only high-order spatial derivative terms in the Lagrangian, while still keeping the time derivative terms to the second order.
This is exactly what Ho\v rava did recently \cite{hovrava2009quantum,wang2017hovrava}.
In addition, another recipe to treat the ghosts is not removing them from the action, while focusing on the higher-order instability in the equations of motion \cite{carroll2005cosmology}.
For the general second-order Lagrangian with propagating torsion, a systematical way to remove the ghosts and tachyons was introduced in \cite{neville1980gravity,sezgin1980new} using spin projection operators.
The gauge fields $(e,\omega)$ can be decomposed irreducibly by $su(2)$ group into different spin modes by means of the weak-field approximation.
In addition to the graviton, three classes spin-$0^{\pm},1^{\pm},2^{\pm}$ modes of torsion were introduced.
\cite{sezgin1980new} studied the general quadratic Lagrangian with nine-parameter and obtained the conditions on the parameters for not having ghosts and tachyons at the massive and massless sectors, respectively.
In this work, to develop a good cosmology based on PGG, we will adopt their nine-parameter Lagrangian with ghost- and tachyons-free conditions on parameters.
The Hamiltonian analysis of PGG for different modes can be found in \cite{yo1999hamiltonian,yo2002hamiltonian}, which tell us that the only safe modes of torsion are spin-$0^{\pm}$, corresponding to the scalar and pseudo-scalar components of torsion, respectively.

It's natural to apply the corresponding Poincar\'e gague cosmology (PGC) on understanding the evolution of the Universe.
The last decade, a series of work \cite{minkevich2007regular,shie2008torsion,minkevich2009accelerating,chen2009cosmological,li2009torsion,baekler2011poincare,ao2010analytical,ao2012torsion,garkun2011numerical,minkevich2013some,ho2015general} (from both analytical and numerical approach) proved that it is possible to reproduce the late-time acceleration in PGC without ``dark energy''.
In Ref. \cite{minkevich2006analysis}, the authors discussed the early-time behaviors of the expanding solution of PGC with a scalar field (inflaton), while in Ref. \cite{wang2009inflation}, a power-law inflation was studied in a $R+R^2$ model of PGC without inflaton.

The current work is a continuation of our previous one: \emph{Late-time acceleration and inflation in a Poincar\'e gauge cosmological model} \cite{zhang2019late}.
In our previous work, we proposed several fundamental assumptions to define the PGC on FLRW level.
Then we studied the general nine-parameter PGC Lagrangian with ghost- and tachyon-free constraints on parameters.
With specific choice of parameters, we obtained two Friedmann-like analytical solutions by varying the Lagrangian, where the scalar torsion $h$-determined solution is consistent with the Starobinsky cosmology in the early time and the $f$-determined solution contains naturally a constant geometric ``dark energy'' density, which cover the $\Lambda$CDM model in the late-time.
We further constrained the magnitudes of parameters using the latest observations.
However, we left a problem unsolved that two solutions are mutually exclusive even they are derived from a same Lagrangian.
The reason comes probably from that the restraint on $B_1$ (vanishing) is too strong, so that the high-order terms of $f$ are removed.
Therefore, in current work, we will investigate the general case at least ghost-free, and use the new results for leading to the slow-roll inflation without any extra fields.
This series of work aims to build self-consistent cosmology to solve the problem of SM in describing the evolution of the Universe, where ``self-consistent'' means without extra hypothesis of inflaton and ``dark energy''.

This paper is organized as follows. In Sec. \ref{cosmological_equations}, we start from the nine-parameter Lagrangian with the ghost- and tachyon-free conditions on parameters.
Then replacing the scalar and pseudo-scalar torsion by two new variables, we rewrite the cosmological equations obtained in \cite{zhang2019late} into new forms.
In Sec. \ref{hybrid_inflation}, we discuss the possibility of the hybrid inflation with a first-order phase transition generated in this gravitational system.
In Sec. \ref{slow-roll_inflation}, we study the slow-roll inflation of this system.
We conclude and discuss our work in Sec. \ref{conclusion}.

To learn more about what the current work based on, please see \cite{zhang2019late} and references therein.

\section{Cosmological equations}\label{cosmological_equations}
We consider the nine-parameter gravitational Lagrangian $\mathcal{L}_G$, which reads:
\begin{alignat}{2}\label{general_action}
I=&\int d^4x\sqrt{\vert g\vert}\big[\frac{1}{2\kappa}\mathcal{L}_G+\mathcal{L}_M\big],\nonumber\\
\mathcal{L}_G=&\alpha R+\mathcal{L}_T+\mathcal{L}_R,\nonumber\\
\mathcal{L}_T\equiv& a_1T_{\mu\nu\rho}T^{\mu\nu\rho}+a_2T_{\mu\nu\rho}T^{\nu\mu\rho}+a_3T_{\mu}T^{\mu},\nonumber\\
\mathcal{L}_R\equiv& b_1R_{\mu\nu\rho\sigma}R^{\mu\nu\rho\sigma}+b_2R_{\mu\nu\rho\sigma}R^{\rho\sigma\mu\nu}+b_3R_{\mu\nu}R^{\mu\nu}\nonumber\\
&+b_4R_{\mu\nu}R^{\nu\mu}+b_5R_{\mu\nu\rho\sigma}R^{\mu\rho\nu\sigma},
\end{alignat}
where $\kappa\equiv8\pi G=8\pi m_{PI}^{-2}$ with $m_{PI}$ the Planck mass, and $\alpha$, $a_1\sim a_3$ are freely dimensionless Lagrangian parameters, while $b_1\sim b_5$ are free Lagrangian parameters with dimension $m_{PI}^{-2}$.
The $R^2$ term need not be included due to the use of the Chern-Gauss-Bonnet theorem \cite{chern1944simple}:
\begin{equation}
\int d^4x\sqrt{\vert g\vert}(R_{\mu\nu\rho\sigma}R^{\mu\nu\rho\sigma}-4R_{\mu\nu}R^{\mu\nu}+R^2)=0,
\end{equation}
for spacetime topologically equivalent to flat space.

For the pair of gauge field ($e,\omega$) as the dynamical variables, the field equations are up to $2$nd-order.
However, the gauge fields ($e,\omega$) can be decomposed irreducibly by $su(2)$ group into different spin modes by means of the weak-field approximation.
In addition to the graviton, three classes spin-$0^{\pm},1^{\pm},2^{\pm}$ modes of torsion were introduced.
It is obvious that in such a general quadratic, the ghosts and tachyons are inevitable for certain modes.
Fortunately, the authors studied this Lagrangian in \cite{sezgin1980new} using the spin projection operators and obtained the conditions on parameters for not having ghosts and tachyons at the massive and massless sectors, respectively.

According to \cite{sezgin1980new}, we summarize the ghost- and tachyon-free conditions on parameters for action (\ref{general_action}) in TABLE I:
\begin{table}[!htbp]
	\caption{The ghost- and tachyon-free conditions on parameters for six spin modes, respectively.}
	\begin{tabularx}{0.48\textwidth}{c|l}
		\hline\hline
		spin modes & conditions on parameters\\
		\hline
		\multirow{2}{*}{$2^-$}&$4 b_1+b_5<0$\\ &$\alpha +2a_1+a_2<0$\\
		\hline
		\multirow{3}{*}{$1^-$}&$4 b_1+2 b_3+b_5<0$\\ &$(\alpha +2a_1+a_2) (2a_1+a_2+a_3)\cdot$\\&$(-2\alpha +2 a_1+a_2+3 a_3)<0$\\
		\hline
		\multirow{2}{*}{$0^-$}&$-2b_1+b_5>0$\\ &$\alpha-4a_1 +4 a_2>0$\\
		\hline
		\multirow{2}{*}{$2^+$}&$4 b_1+4b_2+b_3+b_4+2b_5>0$\\ &$\alpha  (2a_1+a_2) (\alpha +2a_1+a_2)<0$\\
		\hline
		\multirow{2}{*}{$1^+$}&$-4b_1+4 b_2-b_3+b_4<0$\\ &$(2a_1-a_2) (-\alpha +4a_1-4 a_2)(\alpha +2a_1+a_2)>0$\\
		\hline
		\multirow{2}{*}{$0^+$}&$b_1+b_2+b_3+b_4+b_5/2>0$\\ &$\alpha (2 a_1+a_2+3a_3)(-2 \alpha +2a_1+a_2+3a_3)>0$\\
		\hline\hline
	\end{tabularx}
\end{table}

For the massless sector, the ghost-free condition is just: $\alpha>0$.

We will still focus on the FLRW cosmology, where the spatial curvature free metric and non-vanishing components of torsion are, respectively
\begin{equation}
ds^{2} = -dt^{2}+a^{2}d\mathbf{x}^{2},
\label{eq:FRWmetric}
\end{equation}
\begin{equation}
T_{ij0} = a^2 h \delta_{ij}, ~~ T_{ijk} = a^3 f \epsilon_{ijk}, ~~ i,j,k=1,2,3.
\label{eq:TorsionScalar}
\end{equation}
Where $a$, $h$, $f$ are scale factor, scalar torsion, and pseudo-scalar torsion respectively, and are functions of cosmic time $t$.
The general cosmological equations corresponding to the action (\ref{general_action}), on the background can be found in our former work \cite{zhang2019late}, which are cumbersome and the physical meaning lost.
While, we noticed that the equations (23) and (24) in \cite{zhang2019late} can be regarded as the dynamic evolutions of scalar torsion $h$ and pseudo-scalar torsion $f$, and they are second-order of $h$ and $f$, respectively.
We would like to recast them into the Klein-Gordon-like form by introducing the following new variables:
\begin{eqnarray}\label{trans_h_f}
	h&\mapsto&\phi_h=a(h-H),\nonumber\\
	f&\mapsto&\phi_f=af,
\end{eqnarray}
(the scale factors $a$ in transformations are for absorbing the Hubble rate $H$ would occur in the potential to avoid tricky recursions)
then, the cosmological equations (14)$\sim$(17) in \cite{zhang2019late} can be rewritten as:
\begin{widetext}
	\begin{alignat}{4}
	H^2&=\frac{1}{3}\kappa\rho+\frac{1}{3}\kappa\rho_{\phi},\label{eq_friedmann}\\
	2\dot{H}+3H^2&=-\kappa p-\kappa p_{\phi},\\
	\dot{\rho_{\phi}}&=-3H(\rho_{\phi}+p_{\phi}),\\
	\kappa\rho_{\phi}&=\frac{1}{2}\frac{12B_0}{a^2}\dot{\phi_h}^2-\frac{1}{2}\frac{12B_1}{a^2}\dot{\phi_f}^2+V(\phi_h,\phi_f)-3\frac{A_1-2}{2}H^2,\label{eq_krho}\\
	\frac{12B_0}{a}\ddot{\phi_h}&+\frac{12B_0}{a}H\dot{\phi_h}+12(2B_0-2B_1-B_2)\frac{\phi_f}{a^2}\dot{\phi_f}+a\frac{\partial V(\phi_h,\phi_f)}{\partial\phi_h}+3(A_1-2\alpha)H=0,\label{eq_phi_h}\\
	-\frac{12B_1}{a}\ddot{\phi_f}&-\frac{12B_1}{a}H\dot{\phi_f}-12(2B_0-2B_1-B_2)\frac{\phi_f}{a^2}\dot{\phi_h}+a\frac{\partial V(\phi_h,\phi_f)}{\partial\phi_f}=0,\label{eq_phi_f}\\
	V(\phi_h,\phi_f)&=3\frac{A_1-2\alpha}{2}\frac{\phi_h^2}{a^2}-3(4A_0-\alpha)\frac{\phi_f^2}{a^2}-6\frac{B_0}{a^4}(\phi_f^4-2\frac{B_0+2B_1}{B_0}\phi_f^2\phi_h^2+\phi_h^4),\label{eq_potential}
	\end{alignat}
\end{widetext}
where the combinations of parameters are (corresponding to (25) in \cite{zhang2019late}):
\begin{alignat}{2}
A_0&\equiv a_1-a_2,~~A_1\equiv 2a_1+a_2+3a_3,\nonumber\\
B_0&\equiv b_1+b_2+b_3+b_4+\frac{1}{2}b_5,\nonumber\\
B_1&\equiv b_1-\frac{1}{2}b_5,~~B_2\equiv 4b_2+b_3+b_4+b_5.
\end{alignat}
The degeneracy among these Lagrangian parameters on background makes the inequalities can not be solved completely.
However, it is obvious that ghost- and tachyon-free spin-$0^{\pm}$ ``particles'' require:
\begin{equation}\label{ghost-free-spin0}
B_0>0,~~B_1<0,~~\alpha-4A_0>0,~~\alpha A_1(A_1-\alpha)>0.
\end{equation}

Now, the physical picture is quite clear, that the nine-parameter PGC system is equivalent to a gravitational system coupled two-scalar fields $(\phi_h,\phi_f)$, with a potential up to quartic-order, $V(\phi_h,\phi_f)$.
(\ref{eq_phi_h}) and (\ref{eq_phi_f}) are the equations of motion for $\phi_h$ and $\phi_f$, respectively.
They look very symmetrical except the last term in $(\ref{eq_phi_h})$.
If $B_0$, $B_1$ and $(2B_0-2B_1-B_2)$ don't vanish, $|(2B_0-2B_1-B_2)\phi_f/a|$ represents the strength of interaction between two scalar fields.
We conclude that $B_0$ and $B_1$ must have the opposite sign so that the interaction terms in the equations of motion have opposite sign too.
The different between $A_1$ and $\alpha$ measures the weight of $\phi_h^2$ in the potential $V(\phi_h,\phi_f)$, and analogously, $A_0$ and $\alpha$ for $\phi_f^2$.
It will be convenient to overlook the $1/a$ factor in front of field $\phi_h$ or $\phi_f$ because of the inverse factor occurred in (\ref{trans_h_f}).
The ghost- and tachyon-free conditions for spin-$0^{\pm}$ ensure that the kinetic energy terms are positive in (\ref{eq_krho}), as well as a potential well can be formed from (\ref{eq_potential}) when one require $\alpha>0$ and $A_1>0$.
In addition, we notice that when setting $A_1=0$, the last term in (\ref{eq_krho}) will offset $H^2$ in the left hand side of Friedmann equation (\ref{eq_friedmann}), which degrades the entire system into the trivial situations as we studied in our former work \cite{zhang2019late}.
If $A_1\neq0$, to remove the possible recursion in (\ref{eq_krho}), we should set exactly $A_1=2$.

The above system is general because we didn't set any additional assumptions on the parameters yet except the ghost- and tachyon-free conditions (\ref{ghost-free-spin0}).
In the rest of this work, we will focus on the inflationary period of this system, thus the energy densities $\rho$ and pressures $p$ of the matters (with equation of state parameters, i.e. EOS: $w=0,1/3$) will be neglected.
According to the choosing of parameters, the potential (\ref{eq_potential}) can be classified into several types of inflation.

\section{Hybrid inflation with first-order phase transition}\label{hybrid_inflation}
We start from considering two types of hybrid inflation by means of the features of two scalar fields,
\begin{eqnarray}
\text{Case I:}\\
	V_I(\phi_h,\phi_f)&=&[12(B_0+2B_1)\frac{\phi_h^2}{a^2}-3(4A_0-\alpha)]\frac{\phi_f^2}{a^2}-6B_0\frac{\phi_f^4}{a^4}\nonumber\\
	&+&V_I^{eff}(\phi_h),\\
	V_I^{eff}(\phi_h)&\equiv&3(1-\alpha)\frac{\phi_h^2}{a^2}-6B_0\frac{\phi_h^4}{a^4},
\end{eqnarray}
\begin{eqnarray}
\text{Case II:}\\
	V_{II}(\phi_h,\phi_f)&=&[12(B_0+2B_1)\frac{\phi_f^2}{a^2}+3(1-\alpha)]\frac{\phi_h^2}{a^2}-6B_0\frac{\phi_h^4}{a^4}\nonumber\\
	&+&V_{II}^{eff}(\phi_f),\\
	V_{II}^{eff}(\phi_f)&\equiv&-3(4A_0-\alpha)\frac{\phi_f^2}{a^2}-6B_0\frac{\phi_f^4}{a^4},
\end{eqnarray}
where we regard $\phi_h$ as the inflaton caused the slow-roll inflationary phase, and $\phi_f$ as an auxiliary field which can trigger a phase transition, occurring either before or just after the breaking of slow-roll conditions in Case I, and interchange the positions of $\phi_h$ and $\phi_f$ in Case II.

In the standard hybrid inflation, the curvature of potential in the auxiliary direction should be much greater than in the inflaton direction, so that the slow-roll inflation could happen when the auxiliary field rolled down to its minimum value, whereas the inflaton could remain large for a much longer time \cite{linde1994hybrid}.
In Case I, the effective mass squared of the field $\phi_f$ is equal to $2[12(B_0+2B_1)\frac{\phi_h^2}{a^2}-3(4A_0-\alpha)]$.
Therefore for $\phi_h>\phi^c_{h}\equiv\frac{a}{2}\sqrt{\frac{4A_0-\alpha}{B_0+2B_1}}$, the only minimum of the potential is at $\phi_f=0$.
For this reason, we will consider the stage of inflation at large $\phi_h$ with $\phi_f=0$.
However, we notice that in the potential, both $\phi_h$ and $\phi_f$ have the same orders and the same coefficient $-6B_0$ in the quartic term.
$\phi_h$ and $\phi_f$ are highly symmetrical, especially when $A_0$ approximates to $1/4$, so it's difficult to distinguish them from their speed of slow-rolling.
Fortunately, the extra term $6(1-\alpha)H$ in the equation of motion of $\phi_h$ (\ref{eq_phi_h}) may help us to break the symmetry in potential.
During the inflation, the Hubble rate $H$ is approximated as a constant, so it can be absorbed into the potential, and the effective potential of inflton $\phi_h$ in Case I can be modified as
\begin{equation}\label{Veffp}
	V_I^{eff'}(\phi_h)= 6(1-\alpha)H\frac{\phi_h}{a}+3(1-\alpha)\frac{\phi_h^2}{a^2}-6B_0\frac{\phi_h^4}{a^4}.
\end{equation}

When $\phi_h$ falls down to the critical point $\phi^c_h$, the phase transition with another symmetry breaking occurs.
To fulfill the condition of the standard hybrid inflation with small quartic term \cite{lesgourgues2006inflationary}, we assume the $1$st-order term dominates the effective potential of $\phi_h$ in (\ref{Veffp}) at the moment of phase transition (also throughout the inflation), i.e. (substituting $\phi^c_h$ into (\ref{Veffp}))
\begin{eqnarray}\label{condition_hybrid}
	&&\frac{1-\alpha}{4}\frac{4A_0-\alpha}{B_0+2B_1}-\frac{B_0}{8}(\frac{4A_0-\alpha}{B_0+2B_1})^2\nonumber\\
	&\ll& (1-\alpha)H\sqrt{\frac{4A_0-\alpha}{B_0+2B_1}},
\end{eqnarray}
which requires that $4A_0-\alpha\ll B_0+2B_1$.
Then, the Hubble rate $H$ can be estimated as:
\begin{equation}\label{Hubble_c}
	H\cong\kappa (1-\alpha)\sqrt{\frac{4A_0-\alpha}{B_0+2B_1}},
\end{equation}
which is a constant as we expected.
However, it's easy to check that by substituting (\ref{Hubble_c}) into (\ref{condition_hybrid}), the magnitudes on both sides of ``$\ll$'' are at the same order.
This contradicts our assumption of the hybrid inflation with small quartic term, which means there is no a minimum effective potential with a non-zero vacuum energy.
Theoretically speaking, the approach to hybrid inflation with a first-order phase transition failed for Case I.
The same conclusion can be get for Case II.

\section{Slow-roll inflation and numerical analysis}\label{slow-roll_inflation}
According to our presupposition that $B_0>0$, if the coefficients of the quadratic terms in potential (\ref{eq_potential}) are positive with the same magnitude, and $|B_1|$ has the same magnitude with $B_0$, we can get a potential well with a effective radius $r_\phi\lesssim\sqrt{\frac{A_1-2\alpha}{4B_0}}$.
By choosing special values of parameters, the slow-roll inflation can be obtained in this potential well.
In this scenario, the numerical analysis is more clear and convincing than the theoretical analysis.

\begin{figure}[htbp]
	\centering
	\includegraphics[scale=0.9]{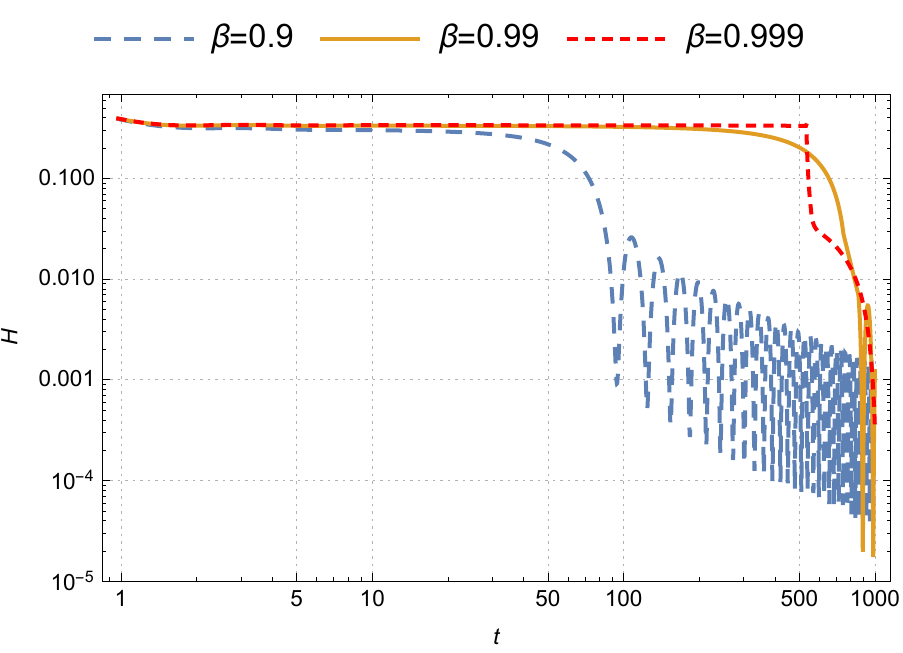}
	\caption{Evolution of Hubble rate $H$ over time. The large dashed (blue), the solid (orange) and the small dashed (red) lines correspond to $\beta=0.9,0.99,0.999$, respectively. To compare with the fiducial value, the smaller $\beta$ makes the decay advanced. The e-folds for three scenarios read $N_{inf}=18.2,160.8,33.1$.}
	\label{fig:Hubble_beta}
\end{figure}

\begin{figure}[htbp]
	\centering
	\includegraphics[scale=0.9]{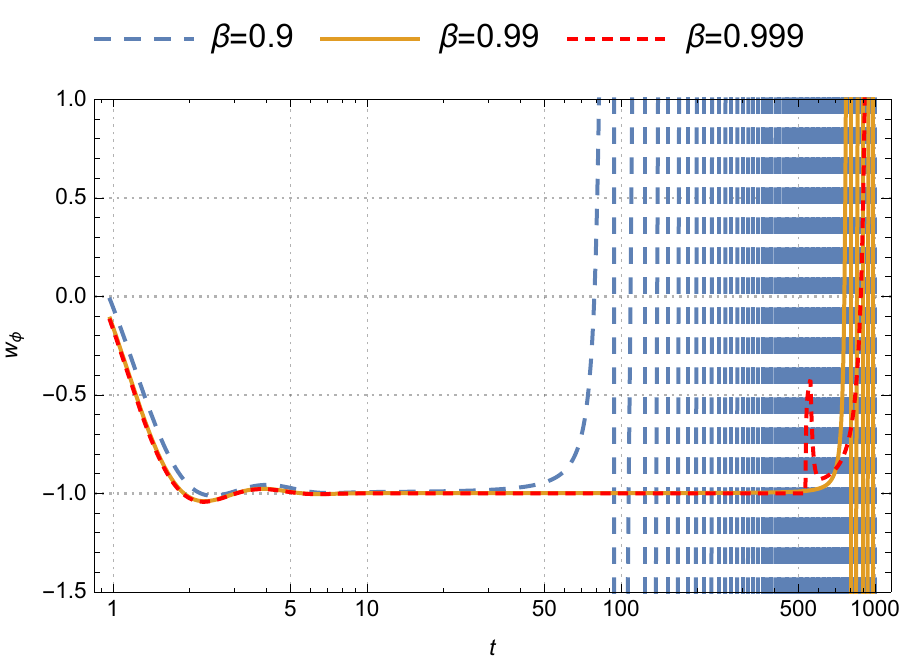}
	\caption{Evolution of EOS $w_{\phi}$ over time. The large dashed (blue), the solid (orange) and the small dashed (red) lines correspond to $\beta=0.9,0.99,0.999$, respectively. $w_{\phi}<-1/3$ during the whole inflationary stage, and $w_{\phi}$ is approximately equal to $-1$ in the deep inflationary stage. The inflation ends when $w_{\phi}=-1/3$ again, then reheating starts. The e-folds for three scenarios read $N_{inf}=18.2,160.8,33.1$.}
	\label{fig:w_beta}
\end{figure}

\begin{figure}[htbp]
	\centering
	\includegraphics[scale=0.9]{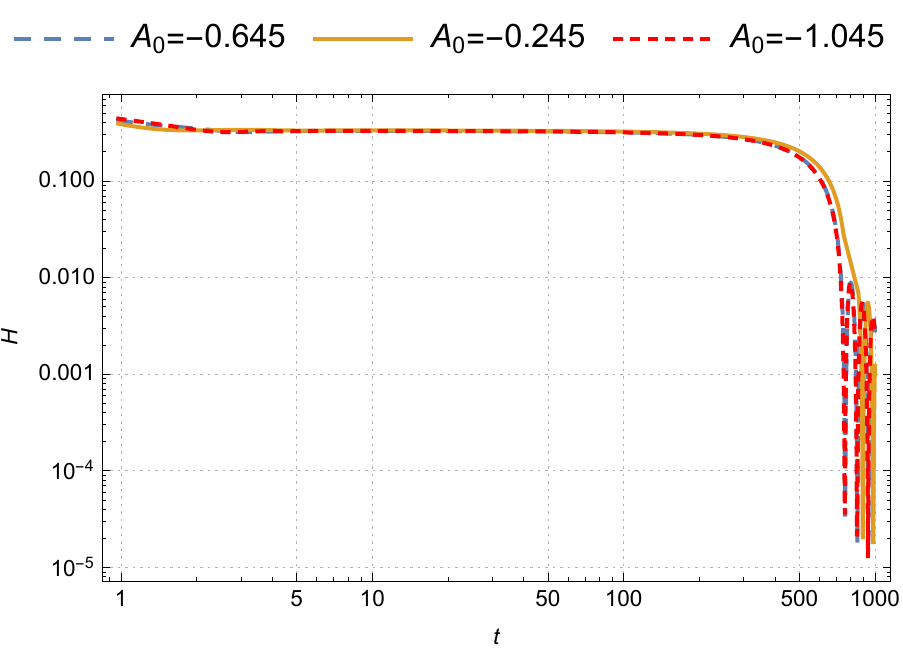}
	\caption{Evolution of Hubble rate $H$ over time. The large dashed (blue), the solid (orange) and the small dashed (red) lines correspond to $A_0=-0.645,-0.245,-1.045$, respectively. We can see that this parameter has little effect on the Hubble rate. The e-folds for three scenarios read $N_{inf}=151.7,160.8,151.1$.}
	\label{fig:Hubble_A0}
\end{figure}

\begin{figure}[htbp]
	\centering
	\includegraphics[scale=0.9]{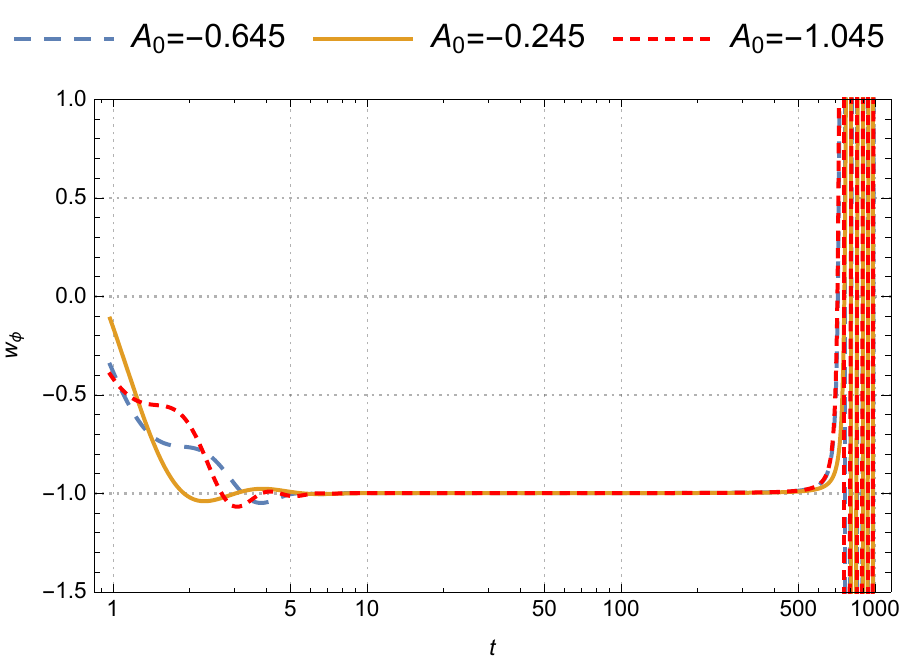}
	\caption{Evolution of EOS $w_{\phi}$ over time. The large dashed (blue), the solid (orange) and the small dashed (red) lines correspond to $A_0=-0.645,-0.245,-1.045$, respectively. $w_{\phi}<-1/3$ during the whole inflationary stage, and $w_{\phi}$ is approximately equal to $-1$ in the deep inflationary stage. The inflation ends when $w_{\phi}=-1/3$ again, then reheating starts. The smaller value of $A_0$ don't influence the decay and reheating but leads to oscillation before inflation. The e-folds for three scenarios read $N_{inf}=151.7,160.8,151.1$.}
	\label{fig:w_A0}
\end{figure}

\begin{figure}[htbp]
	\centering
	\includegraphics[scale=0.9]{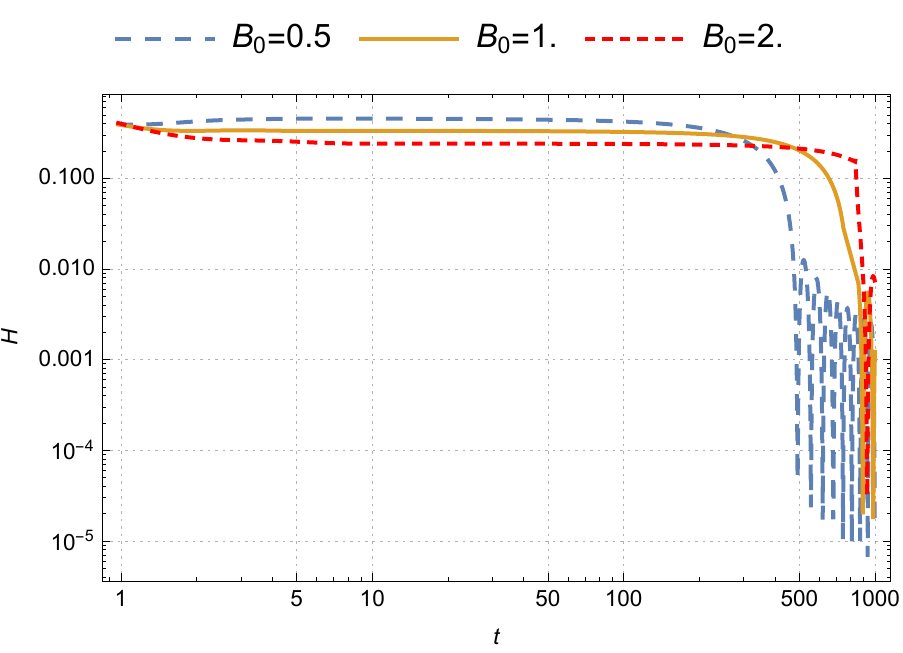}
	\caption{Evolution of Hubble rate $H$ over time. The large dashed (blue), the solid (orange) and the small dashed (red) lines correspond to $B_0=0.5,1.0,2.0$, respectively. To compare with the fiducial value, the smaller $B_0$ makes the decay advanced. The e-folds for three scenarios read $N_{inf}=139.2,160.8,82.2$.}
	\label{fig:Hubble_B0}
\end{figure}

\begin{figure}[htbp]
	\centering
	\includegraphics[scale=0.9]{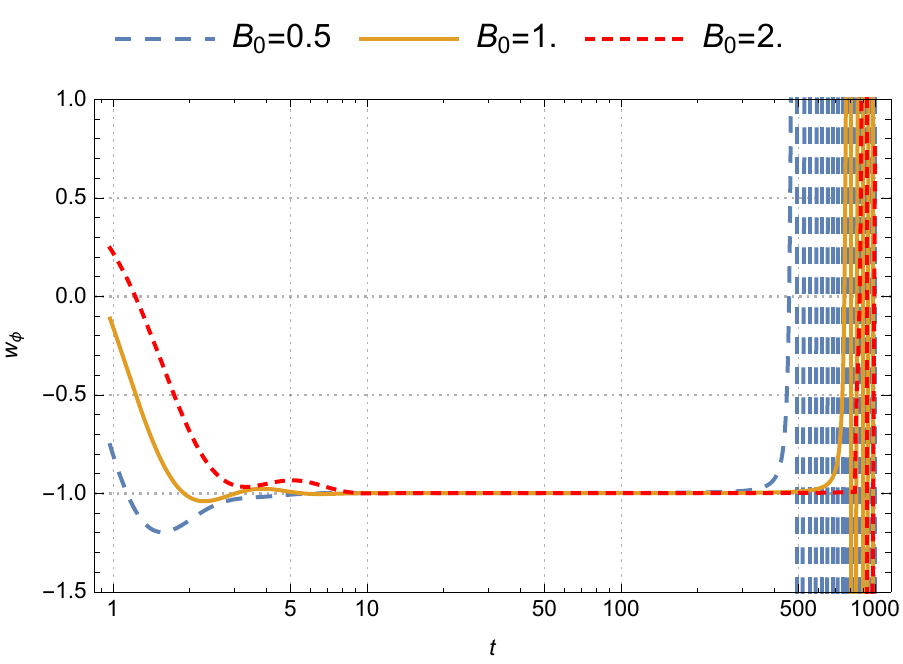}
	\caption{Evolution of EOS $w_{\phi}$ over time. The large dashed (blue), the solid (orange) and the small dashed (red) lines correspond to $B_0=0.5,1.0,2.0$, respectively. $w_{\phi}<-1/3$ during the whole inflationary stage, and $w_{\phi}$ is approximately equal to $-1$ in the deep inflationary stage. The inflation ends when $w_{\phi}=-1/3$ again, then reheating starts. To compare with the fiducial value, the smaller $B_0$ makes the curve overall left shift, while right shift for larger $B_0$. The e-folds for three scenarios read $N_{inf}=139.2,160.8,82.2$.}
	\label{fig:w_B0}
\end{figure}

\begin{figure}[htbp]
	\centering
	\includegraphics[scale=0.9]{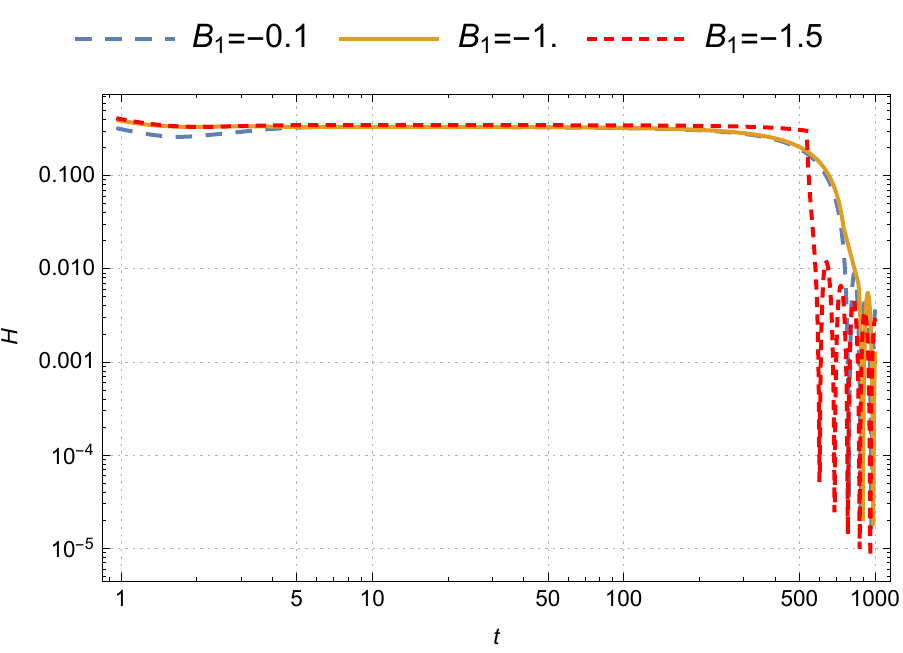}
	\caption{Evolution of Hubble rate $H$ over time. The large dashed (blue), the solid (orange) and the small dashed (red) lines correspond to $B_1=-0.1,-1.0,-1.5$, respectively. To compare with the fiducial value, the larger absolute value of $B_1$ makes the decay advanced. The e-folds for three scenarios read $N_{inf}=157.2,160.8,55.6$.}
	\label{fig:Hubble_B1}
\end{figure}

\begin{figure}[htbp]
	\centering
	\includegraphics[scale=0.9]{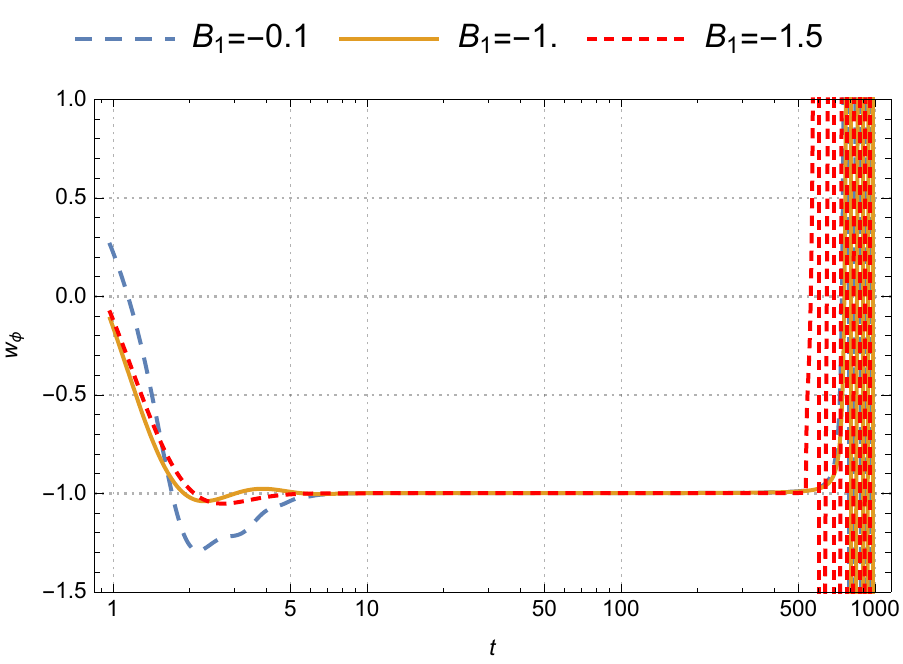}
	\caption{Evolution of EOS $w_{\phi}$ over time. The large dashed (blue), the solid (orange) and the small dashed (red) lines correspond to $B_1=-0.1,-1.0,-1.5$, respectively. $w_{\phi}<-1/3$ during the whole inflationary stage, and $w_{\phi}$ is approximately equal to $-1$ in the deep inflationary stage. The inflation ends when $w_{\phi}=-1/3$ again, then reheating starts. The larger absolute value of $B_1$ makes the decay advanced, while the smaller absolute value of $B_1$ makes $w_{\phi}$ sinking before inflation. The e-folds for three scenarios read $N_{inf}=157.2,160.8,55.6$.}
	\label{fig:w_B1}
\end{figure}

\begin{figure}[htbp]
	\centering
	\includegraphics[scale=0.9]{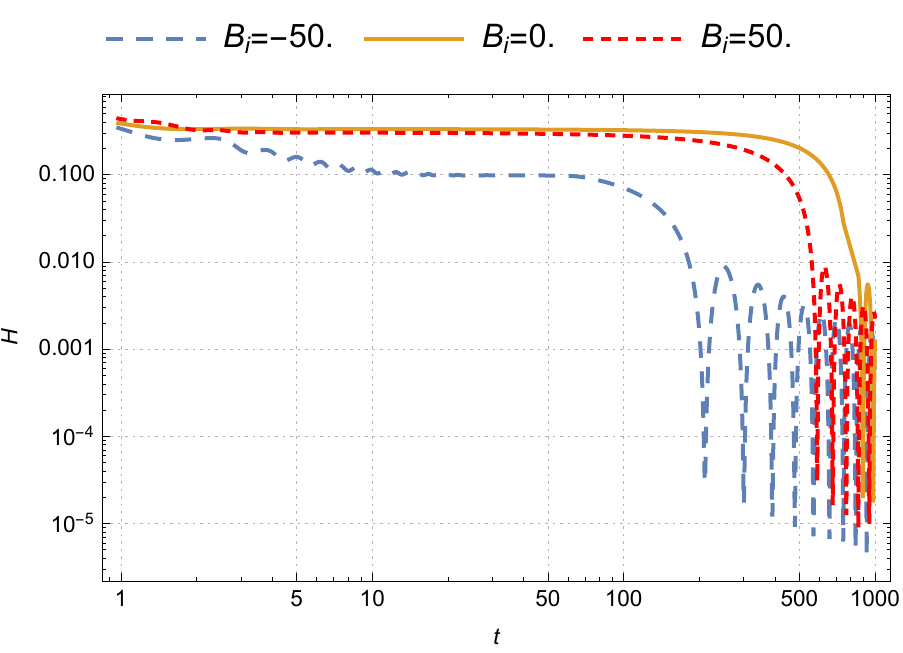}
	\caption{Evolution of Hubble rate $H$ over time. The large dashed (blue), the solid (orange) and the small dashed (red) lines correspond to $B_i=-50,0,50$, respectively. To compare with the fiducial value, any interaction between two-scalar fields makes the decay advanced. The e-folds for three scenarios read $N_{inf}=12.6,160.8,105.6$.}
	\label{fig:Hubble_Bi}
\end{figure}

\begin{figure}[htbp]
	\centering
	\includegraphics[scale=0.9]{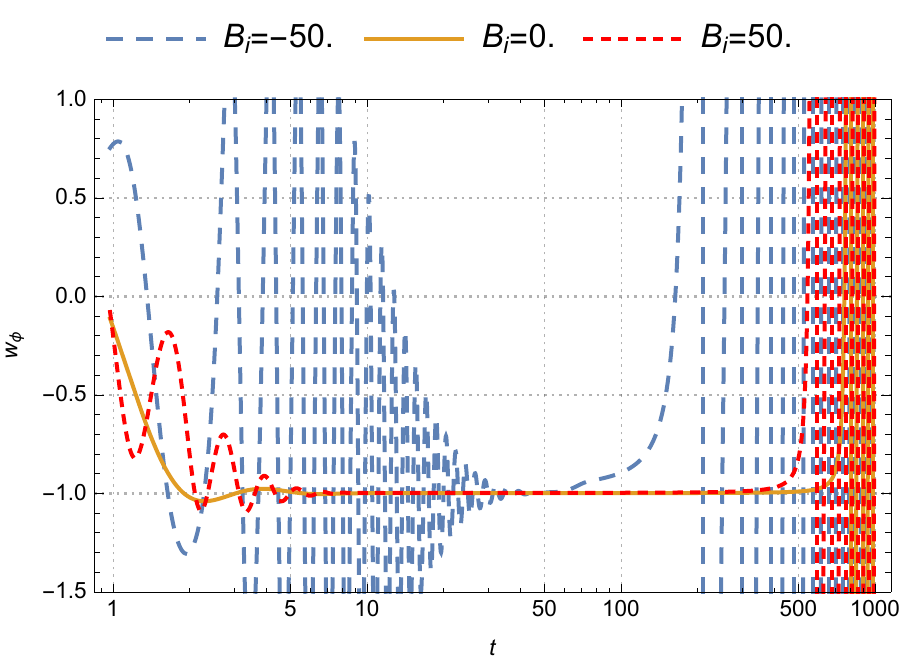}
	\caption{Evolution of EOS $w_{\phi}$ over time. The large dashed (blue), the solid (orange) and the small dashed (red) lines correspond to $B_i=-50,0,50$, respectively. $w_{\phi}<-1/3$ during the whole inflationary stage, and $w_{\phi}$ is approximately equal to $-1$ in the deep inflationary stage. The inflation ends when $w_{\phi}=-1/3$ again, then reheating starts. The interaction between two-scalar fields leads to oscillation on $w_{\phi}$ before inflation. The e-folds for three scenarios read $N_{inf}=12.6,160.8,105.6$.}
	\label{fig:w_Bi}
\end{figure}

The dimensions for parameters and quantities read:
\begin{alignat}{2}
	\alpha\sim A_0\sim A_1&\sim1,\nonumber\\
	B_0\sim B_1\sim B_2&\sim m_{PI}^{-2},\nonumber\\
	t&\sim m_{PI}^{-1},\nonumber\\
	H\sim\phi_h\sim\phi_f&\sim m_{PI},\nonumber\\
	\dot{\phi_h}\sim\dot{\phi_f}&\sim m_{PI}^{2}.
\end{alignat}
In the unit of $m_{PI}=1$, and by considering that $a$ can be rescaled, we set the initial data (labeled with ``B'') as:
\begin{alignat}{2}\label{initial_data}
	t_B=0,~~a_B&=1,\nonumber\\
	\phi_h(t_B)=\phi_f(t_B)&=0,\nonumber\\
	\dot{\phi_h}(t_B)=\dot{\phi_f}(t_B)&=1.
\end{alignat}
To investigate the effect of each parameter on this system, we choose a set of fiducial values:
\begin{alignat}{2}\label{fiducial_values}
	\beta&\equiv 1-\alpha=0.99,~~A_0=-0.245,\nonumber\\
	B_0&=1,~~B_1=-1,\nonumber\\
	B_i&\equiv 2B_0-2B_1-B_2=0,
\end{alignat}
then vary a parameter and plot while keeping other parameters to maintain the fiducial value.
FIG. \ref{fig:Hubble_beta}, \ref{fig:Hubble_A0}, \ref{fig:Hubble_B0}, \ref{fig:Hubble_B1}, \ref{fig:Hubble_Bi} are the evolution curves of Hubble rate $H$ for every choice of parameter, while FIG. \ref{fig:w_beta}, \ref{fig:w_A0}, \ref{fig:w_B0}, \ref{fig:w_B1}, \ref{fig:w_Bi} for the EOS of scalar fields $w_{\phi}$, where $w_{\phi}\equiv p_{\phi}/\rho_{\phi}$.

These evolution curves show that the system starts from a ``pre-inflation'' stage, then enters into the slow-roll inflation, meanwhile the EOS changes from positive to negative.
After a nearly constant stage, the system decay rapidly, which we call it ``pre-reheating''.
Then the subsequent oscillation indicates that the system has entered a stage of reheating.
To match the current observations, the stage of inflation should last long enough, which can be quantified by e-folds $N_{inf}$:
\begin{equation}
N_{inf}:=\ln\frac{a_f}{a_i},
\end{equation}
where $a_i$ is the scale factor at the moment $t_i$ of the inflationary onset, which is defined by the time when the Universe begins to accelerate $\ddot{a}(t_i)=0$, i.e. $\ddot{a}(t)$ first changes its sign right after the bouncing phase \cite{zhu2017pre}.
The end of the inflation is defined by the time $t_f$ when the accelerating expansion of the Universe stops, i.e. $w_{\phi}(t_f)=-1/3$.
The current observations require that $N_{inf}>60$.

According to the numerical analysis, we summarize the influences of every parameter comparing with the fiducial value as following: 1) smaller $\beta$ can lead to the decay advanced; 2) $A_0$ doesn't influence the decay but causes oscillation before inflation; 3) smaller $B_0$ leads to the curve overall left shift; 4) smaller $|B_1|$ makes $w_{\phi}$ before inflation; 5) the interaction between two-scalar fields makes the decay advanced and leads to oscillation on $w_{\phi}$ before inflation.

\begin{figure}[htbp]
	\centering
	\includegraphics[scale=0.9]{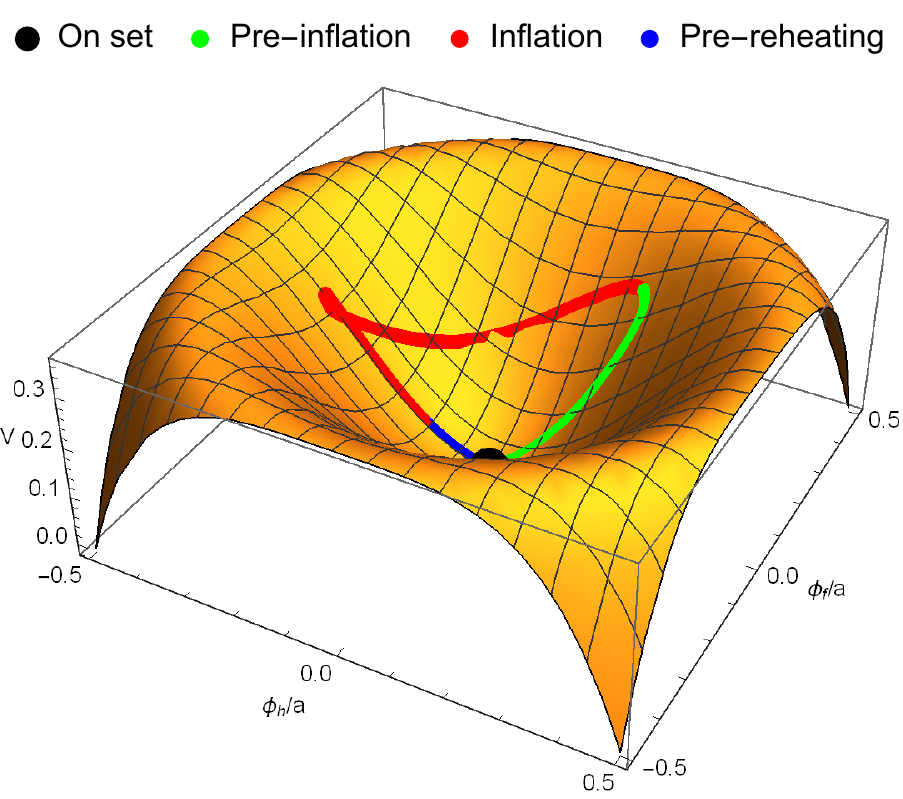}
	\caption{The 3D phase diagram of two-scalar fields $\phi_h$ and $\phi_f$ (over $a$) evolute in the potential well in the fiducial scenario. The shape of potential well is determined by the values of parameters given by (\ref{fiducial_values}). The initial data is given by (\ref{initial_data}) where the initial kinetic energies (speeds) read $\dot{\phi_h}(t_B)=1$, $\dot{\phi_f}(t_B)=1$.}
	\label{fig:v3d}
\end{figure}
\begin{figure}[htbp]
	\centering
	\includegraphics[scale=0.9]{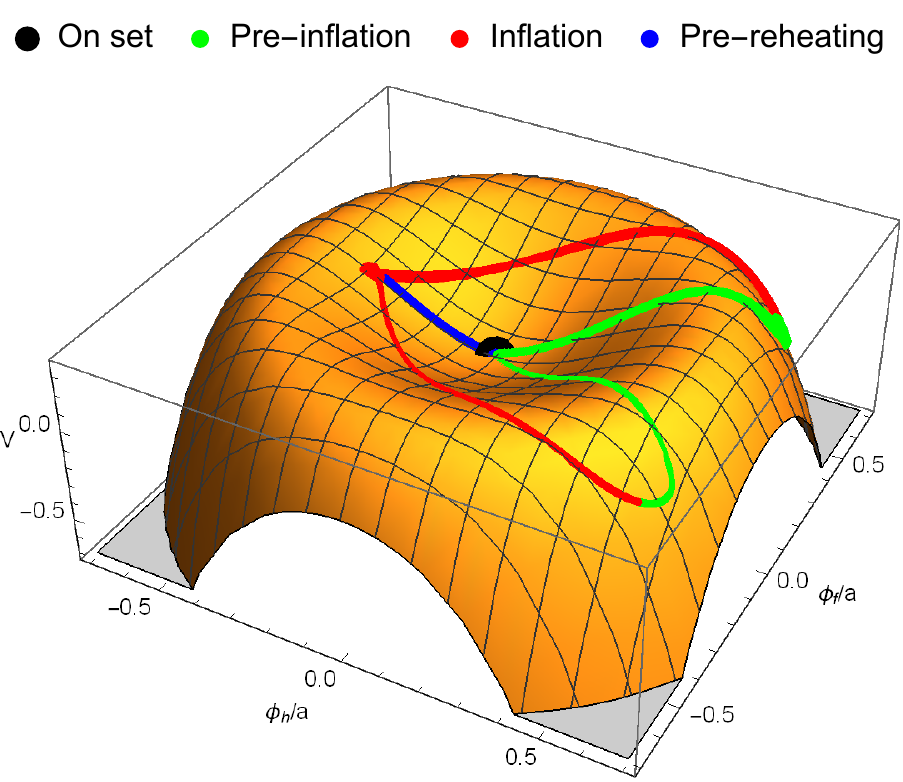}
	\caption{The 3D phase diagram of two-scalar fields $\phi_h$ and $\phi_f$ (over $a$) evolute in the potential well in the fiducial scenario (\ref{fiducial_values}), but start with very large initial kinetic energies (speeds): $\dot{\phi_h}(t_B)=1000$, $\dot{\phi_f}(t_B)=500$ corresponding to the thick trajectory and $\dot{\phi_h}(t_B)=1000$, $\dot{\phi_f}(t_B)=-500$ to the thin one. The potential well looks shallower than the previous one in FIG. \ref{fig:v3d}, not because we changed the fiducial parameters, but instead expanded the ranges of $\phi_h$ and $\phi_f$ (over $a$).}
	\label{fig:v3d_k}
\end{figure}
The 3D phase diagram FIG. \ref{fig:v3d} visualizes the evolutions of two-scalar fields $\phi_h$ and $\phi_f$ (over $a$) on the potential well in the fiducial scenario.
The system starts from the on set point where we set the initial data (\ref{initial_data}), where the non-trivial values are the kinetic energies of two fields.
Then the system is driven by the kinetic energies and climbs to the high level of the potential well, and prepares for the slow-roll inflation.
The slow-roll inflation occurs on the wall of the potential well, where the deep inflation happened after the inflection point, where $\phi_f$ is approximate to $0$, and $\phi_h$ is almost constant.
At last, the system decays and drops down from the wall of the potential well, towards the on set point of the phase space, then leads to the reheating.

It seems that if the initial kinetic energies are large enough, the system may cross the highest point of the potential well, causing it to collapse.
To test the stableness of this system, we keep the fiducial values of parameters (\ref{fiducial_values}) unchanged, but increase the initial kinetic energies (speeds) of two-scalar fields.
FIG. \ref{fig:v3d_k} is the 3D phase diagram of this case, where we set two pairs of very large initial kinetic energies (speeds): $\dot{\phi_h}(t_B)=1000$, $\dot{\phi_f}(t_B)=500$ and $\dot{\phi_h}(t_B)=1000$, $\dot{\phi_f}(t_B)=-500$.
Both trajectories can cross the highest point of the potential well, but can return to the potential well and cause inflation anyway.
Numerical analysis shows that the system has good stability.

\section{Conclusion and discussion}\label{conclusion}
PGG as a gauge field gravitational theory is a natural extension of Einstein's GR to the Poincar\'e group.
It is worth looking forward using PGC, the cosmology of PGG, to solve the problems in the cosmological SM, especially the mechanisms of inflation and late-time acceleration.
In this work, we started from the general nine-parameter gravitational Lagrangian of PGC, and introduced the ghost- and tachyon-free conditions for this Lagrangian.
By introducing new variables $\{\phi_h,\phi_f\}$ for replacing the scalar and pseudo-scalar torsion $\{h,f\}$, we found the general PGC on background is equivalent to a gravitational system coupled to two-scalar fields with a potential up to quartic-order.
We analyzed the possibility of this system producing the hybrid inflation with first-order phase transition, and concluded that it is not feasible.
Then by choosing appropriate parameters, we constructed a potential well from the quartic-order potential, and studied the slow-roll inflation numerically.
We chose a set of fiducial values for parameters, and investigated the effects of each parameter on this system.
All the evolution curves show that this system experiences four different stages: ``pre-inflation'' (on set), slow-roll inflation, ``pre-reheating'' (decay) and reheating.
Most scenarios possess large enough e-folds which is required by the current theories and observations.
The 3D phase diagram of two-scalar fields shows clearly four stages of the evolution in the potential well.
At last, we studied the stableness of this system by setting large values of initial kinetic energies (speeds).
We found that even if the system evolves past the highest point of the potential well, the scalar fields can still return to the potential well and cause inflation.
In short, the numerical analysis for this general PGC system on background indicated that it is a good self-consistent candidate for the slow-roll inflation.
Further studies on the aspect of perturbation will be our next work, especially the primordial power spectrum from this system and it's effects on CMB.
It is also worth looking forward to unify the inflation and the late-time acceleration under PGC in the future.
\begin{acknowledgments}
We thank Prof. Abhay Ashtekar for helpful comments.
Lixin Xu is supported in part by National Natural Science Foundation of China under Grant No. 11675032.
Hongchao Zhang is supported from the program of China Scholarships Council No. 201706060084.
\end{acknowledgments}

\bibliography{PGGC_inflation_1}

\end{document}